\documentclass[final]{aipproc}
\usepackage{amsmath,amssymb}
\layoutstyle{8x11double}

\SetInternalRegister\hbadness{8000} 

\newcommand{\newc}{\newcommand}
\newc\eg{{\it {e.g.}}}  \newc\etal{{\it {et al.}}} \newc\ie{{\it i.e.}}
\newc\etc{{\it {etc}}}  

\newcommand\lsim{\mathrel{\rlap{\lower4pt\hbox{\hskip1pt$\sim$}}
    \raise1pt\hbox{$<$}}}
\newcommand\gsim{\mathrel{\rlap{\lower4pt\hbox{\hskip1pt$\sim$}}
    \raise1pt\hbox{$>$}}}

\newc{\mhalf}{m_{1/2}}      \newc{\mzero}{m_0}

\newc{\tanb}{\tan\beta}
\newc{\azero}{A_0}
\newc{\at}{A_t} \newc{\abot}{A_b} \newc{\atau}{A_\tau} 

\newc{\bmu}{B\mu}           \newc{\sgn}{{\rm sgn}}
\newc{\mone}{M_1}           \newc{\mtwo}{M_2}

\newc{\charone}{\chi_1^\pm} \newc{\mcharone}{m_{\chi_1^\pm}}

\newc{\hl}{h}               \newc{\mhl}{m_{\hl}}
\newc{\hh}{H}               \newc{\mhh}{m_{\hh}}
\newc{\ha}{A}               \newc{\mha}{m_{\ha}}
\newc{\hc}{H^{\pm}}         \newc{\mhc}{m_{\hc}}

\newc{\mw}{m_{W}}      \newc{\mz}{m_{Z}}

\newc{\mgut}{M_{\rm GUT}}

\newc{\mplanck}{M_{\rm P}}      \newc{\mpl}{M_{\rm Pl}}
\newc{\msusy}{M_{\rm SUSY}}      \newc{\ms}{M_{\rm S}}

\newc{\jxf}{J({\xf})}
\newc{\jxfexact}{J_{\rm exact}({\xf})}  \newc{\jxfexp}{J_{\rm exp}({\xf})}
\newc{\VEV}[1]{\langle #1 \rangle}

\newc{\xf}{x_f}
\newc\vrel{v_{\rm rel}}
\newcommand\mchi{m_{\chi}}              

\newc\sell{{\widetilde e}_L}      \newc\msell{m_{\sell}}
\newc\selr{{\widetilde e}_R}      \newc\mselr{m_{\selr}}

\newc\snue{{\widetilde \nu}_e}      \newc\msnue{m_{\snue}}
\newc\snutau{{\widetilde \nu}_\tau}      \newc\msnutau{m_{\snutau}}

\newc\supl{{\widetilde u}_L}      \newc\msupl{m_{\supl}}
\newc\supr{{\widetilde u}_R}      \newc\msupr{m_{\supr}}
\newc\sdl{{\widetilde d}_L}      \newc\msdl{m_{\sdl}}
\newc\sdr{{\widetilde d}_R}      \newc\msdr{m_{\sdr}}

\newcommand{\stauone}{{\tilde \tau}_1}   \newcommand\mstauone{m_{\stauone}}

\newcommand\gluino{\tilde g}
\newcommand\mgluino{m_{\gluino}}

\newc\hpm{H^\pm} \newc\hp{H^+} \newc\hm{H^-} 
\newc\sfermion{\tilde f}  \newc\msfermion{m_{\sfermion}}  

\newc\second{{\rm sec}} 

\newc\alphas{\alpha_s}

\newc\alphaem{\alpha_{em}}

\newc\sigmabar{{\overline{\sigma}}}

\newcommand\treh{T_{\rm R}}
   
\newcommand\Td{T_{\rm D}}

\newc{\sthw}{\sin\theta_W}              \newc{\cthw}{\cos\theta_W}
\newc{\bino}{\widetilde B}              \newc{\wino}{\widetilde W_3}
\newc{\higgsinob}{{\widetilde H}^0_b}   \newc{\higgsinot}{{\widetilde H}^0_t}

\newc{\abund}{\Omega h^2}
\newc{\abundchi}{\Omega_\chi h^2}
\newc{\abundcdm}{\Omega_{{\rm CDM}} h^2}

\newc{\omegam}{\Omega_{{\rm M}}}       \newc{\abundm}{\Omega_{{\rm M}} h^2}
\newc{\omegab}{\Omega_{{\rm b}}}	\newc{\abundb}{\Omega_{{\rm b}} h^2}
\newc{\omegatot}{\Omega_{{\rm TOT}}}

\newc{\omeganlsp}{\Omega_{{\rm NLSP}}}   \newc{\abundnlsp}{\Omega_{\rm NLSP}h^2}
\newc{\ynlsp}{Y_{{\rm NLSP}}}            \newc{\taunlsp}{\tau_{{\rm NLSP}}}
\newc{\nnlsp}{n_{{\rm NLSP}}}            \newc{\mnlsp}{m_{{\rm NLSP}}}
\newc{\nx}{n_{X}}                        \newc{\yx}{Y_{X}}
\newc{\mx}{m_{X}}                        \newc{\taux}{\tau_{X}}

\newc{\rhocrit}{\rho_{crit}}
\newc{\rhochi}{\rho_{\chi}}

\newcommand\fa{f_{a}}

\newcommand\stau{\tilde{\tau}}

\newcommand\neut{\tilde \chi}

\newc{\cachigamma}{C_{a\neut\gamma}}
\newc{\caww}{C_{aWW}}                   
\newc{\cayy}{C_{aYY}}

\newc{\nl}{\cos \theta_{\tilde t}}
\newc{\nr}{\sin \theta_{\tilde t}}
\newcommand\tev{\,\mbox{TeV}}
\newcommand\gev{\,\mbox{GeV}}

\newcommand\kev{\,\mbox{keV}}
\newcommand\ev{\,\mbox{eV}}

\newc\gbar{{\overline{g}}}

\newcommand\cmeter{\,\mbox{cm}}

\newcommand\pb{\,\mbox{pb}}

\newc\snu{{\widetilde \nu}}

\newc{\ra}{\rightarrow}

\newc{\beq}{\begin{equation}}
\newc{\eeq}{\end{equation}}
\newc{\bea}{\begin{eqnarray}}
\newc{\eea}{\end{eqnarray}}

\renewcommand\({\left(}
\renewcommand\){\right)}

\newc{\nspin}{n_{\rm spin}}
\newc{\nflavor}{n_{\rm F}}
\newc{\ngamma}{n_\gamma}
\newc{\ychi}{Y_{\chi}}                  \newc{\yeqchi}{Y^{\rm EQ}_{\chi}}

\newcommand\axino{\tilde{a}}        
\newcommand\maxino{m_{\axino}}

\newcommand\abunda{\Omega_{\axino}h^2} 

\newcommand\abundantp{\Omega^{\rm NTP}_{\axino}h^2}     

\newcommand\abundatp{\Omega^{\rm TP}_{\axino}h^2}       

\newc{\naxino}{n_{\axino}}
\newc{\yaxino}{Y_{\axino}}
\newc{\yeqaxino}{Y^{\rm EQ}_{\axino}}
\newc{\ythaxino}{Y^{\rm TP}_{\axino}}
\newc{\ynthaxino}{Y^{\rm NTP}_{\axino}}

\newcommand\gravitino{\widetilde{G}}    
\newcommand\mgravitino{m_{\gravitino}}

\newcommand\abundg{\Omega_{\gravitino}h^2}

\newcommand\abundgtp{\Omega^{\rm TP}_{\gravitino}h^2}       

\newc{\ngravitino}{n_{\gravitino}}
\newc{\ygravitino}{Y_{\gravitino}}
\newc{\yeqgravitino}{Y^{\rm EQ}_{\gravitino}}
\newc{\ythgravitino}{Y^{\rm TP}_{\gravitino}}
\newc{\ynthgravitino}{Y^{\rm NTP}_{\gravitino}}

\newc\sigmaint{\sigma_{int}}
\newc\sigmaew{\sigma_{\rm weak}}

\newc{\yascat}{Y^{\rm scat}_{i,j}}      \newc{\yadec}{Y^{\rm dec}_{i}}
\newc{\gstar}{g_\ast}           \newc{\gsstar}{g_{s\ast}}

       \def\pslash{\not{\hbox{\kern-2.3pt $p$}}}
       \def\kslash{\not{\hbox{\kern-2.3pt $k$}}}
       \def\qslash{\not{\hbox{\kern-2.3pt $q$}}}
       \def\ddslash{\not{\hbox{\kern-2.3pt $d$}}}
       \def\prtslash{\not{\hbox{\kern-2.3pt $\partial$}}}

\newcommand\apj[3]    {
		{{\it Astrophys.\ J. }{\bf #1} (#2) #3}}

\newcommand\jcap[3] 
		{{\it J.\ Cosmol.\ Astrop.\ Phys.\ }{\bf #1} (#2) #3}
\newcommand\jhep[3]   {
		{{\it J. High Energy Phys.\ }{\bf #1} (#2) #3}}

\newcommand\npb[3]    {
		{{\it Nucl.\ Phys.\ }{\bf B #1} (#2) #3}}
\newcommand\npps[3]   {
		{{\it Nucl.\ Phys.\ }{\bf #1} {\it(Proc.\ Suppl.)} (#2) #3}}

\newcommand\plb[3]    {
		{{\it Phys.\ Lett.\ }{\bf B #1} (#2) #3}}

\newcommand\prd[3]    {
		{{\it Phys.\ Rev.\ }{\bf D #1} (#2) #3}}

\newcommand\prep[3]   {
		{{\it Phys.\ Rept.\ }{\bf #1} (#2) #3}}
\newcommand\prl[3]    {
		{{\it Phys.\ Rev.\ Lett.\ }{\bf #1} (#2) #3}}

\newcommand\sjnp[3]   {
		{{\it Sov.\ J.\ Nucl.\ Phys.\ }{\bf #1} (#2) #3}}

\begin{document}

\title[]{E--WIMPs\footnote{ Invited plenary talk given by
    L. Roszkowski at PASCOS--05, 
Gyeongju, Korea, 30 May -- 4 June 2005.}}


\author{Ki-Young Choi}{
  address={Department of Physics and Astronomy, University of Sheffield, 
Sheffield, S3 7RH, UK}
}

\author{Leszek Roszkowski}{
  address={Department of Physics and Astronomy, University of Sheffield, 
Sheffield, S3 7RH, UK}
}


\begin{abstract}
Extremely weakly interacting massive particles (E--WIMPs) are
intriguing candidates for cold dark matter in the Universe.  We review
two well motivated E--WIMPs, an axino and a gravitino, and point out
their cosmological and phenomenological similarities and differences,
the latter of which may allow one to distinguishing them in LHC searches
for supersymmetry.
\end{abstract}

\date{\today}


\maketitle
\section{Introduction}
From the particle physics point of view, a WIMP (weakly interacting
massive particle) looks rather attractive as a candidate for cold dark
matter (CDM) in the Universe. In many extensions of the Standard Model
(SM) there often exist several new WIMPs, and it is often not too
difficult to ensure that the lightest of them is stable by means of
some discrete symmetry or topological invariant. (For example, in
supersymmetry, one usually invokes $R$--parity.)  In order to meet
stringent astrophysical constraints on exotic relics (\eg, anomalous
nuclei), WIMPs must be electrically and (preferably) color
neutral. They can however interact weakly. For WIMPs produced via a
usual freeze--out from an expanding plasma one finds $\abund\simeq
1/\left\langle\left(\frac{\sigma_{\rm
ann}}{10^{-38}\cmeter^2}\right)\left(\frac{v/c}{0.1}\right)
\right\rangle$. Assuming a pair--annihilation cross section
$\sigma_{\rm ann}\sim\sigma_{\rm weak}\sim 10^{-38}\cmeter^2$, and
since the relative velocity $v$ at freeze--out is non--relativistic, one often
obtains $\abund\sim0.1$, in agreement with current
determinations. This has sometimes been used as a hint for a deeper
connection between weak interactions and CDM in the Universe.

Contrary to this simple and persuasive argument, CDM particles are not
bound to interact with roughly the weak interaction strength.
Extremely weakly interacting massive particles 
(E--WIMPs) have also been known to be excellent candidates for CDM.
In comparison with ``standard'' WIMPs, E--WIMP interaction strength
with ordinary matter is strongly suppressed by some large mass scale,
for example the (reduced) Planck scale $\mplanck\simeq
2.4\times10^{18}\gev$ for gravitino or the Peccei--Quinn scale $f_a\sim
10^{11}\gev$ for axion and/or axino.

E--WIMPs are also well motivated from a particle theorist's
perspective if one takes the point of view that CDM candidates should
appear naturally in some reasonable frameworks beyond the SM which
have been invented to address some other major puzzle in particle
physics. In other words, it would be preferable if a CDM candidate
were not invented for the sole purpose of solving the DM problem.

\begin{figure}[t!]
  \includegraphics[height=.35\textheight]{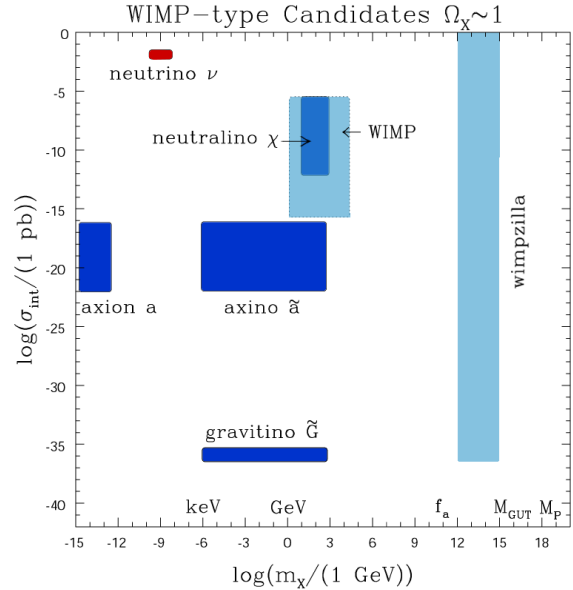}
\caption{A schematic representation of some
well--motivated WIMP--type particles for which a priori one can have
$\Omega\sim1$. $\sigma_{\rm int}$ represents a typical order of
magnitude of interaction strength with ordinary matter. The neutrino
provides hot DM which is disfavored. The box marked ``WIMP' stands for
several possible candidates, \eg, from Kaluza--Klein scenarios.
}
\label{bigpic:fig}
\end{figure}

One way to present well--motivated CDM candidates is to consider a big
``drawing board'' as in 
Fig.~\ref{bigpic:fig}: a plane spanned by the mass of the relic on the
one side and by a typical strength $\sigmaint$ of its interaction with
ordinary matter (\ie, detectors) on the other. To a first
approximation the mass range can in principle extend up to the Planck
mass scale, but not above, if we are talking about elementary
particles. The interaction cross section could reasonably be expected
to be of the weak strength 
($\sigma_{\rm weak} \simeq 10^{-2}\pb$) but could also
be as tiny 
as that purely due to gravity:
$\sim\(\mw/\mplanck\)^2\sigma_{\rm weak}\sim10^{-32} 
\sigma_{\rm weak}\sim10^{-34}\pb$.

What can we put into this vast plane shown in Fig.~\ref{bigpic:fig}?
One obvious candidate is the neutrino, since we know that it exists.
Neutrino oscillation experiments have basically convinced us that its
mass is probably of order $\sim0.1\ev$, or less. On the upper side, if
it were heavier than a few $\ev$, it would overclose the Universe. The
problem is that such a WIMP would constitute {\em hot} DM which is
hardly anybody's favored these days. While some like it hot, or warm,
most like it cold.


The main suspect for today is of course the neutralino
$\chi$~\cite{jkg90,lrpramana03}. While in general it is a mass
eigenstate of a bino, wino and two neutral higgsinos, on the grounds of
naturalness~\cite{lr91} or unification~\cite{kkrw}
it should preferably be mostly a bino. LEP bounds on its
mass are actually not too strong, nor robust: they depend on a number
of assumptions.  In minimal SUSY (the so-called MSSM) ``in most cases''
$\mchi\gsim70\gev$, but the bound can be also much lower.
Theoretically, because of the fine tuning argument, one expects its
mass to lie in the range of several tens or hundreds of
$\gev$~\cite{lr91}. More generally, $\mchi\gsim {\rm few}\gev$ from
$\abundchi\lsim1$ (the so--called Lee-Weinberg
bound~\cite{leeweinberg77}) and $\mchi\lsim340\tev$ from
unitarity~\cite{gk90}. Neutralino interaction rates are generally
suppressed relative to $\sigmaew$ by various mixing angles in the
neutralino couplings. In the MSSM they are typically between
$\sim10^{-3}\sigmaew$ and $\sim10^{-10}\sigmaew$, although could be
even lower in more complicated models where the LSP would be dominated
for example by a singlino component (fermionic partner of an additional Higgs
singlet under the SM gauge group). This uncertainty of the precise
nature of the neutralino is reflected in Fig.~\ref{bigpic:fig} by
showing both a smaller (dark blue) region of minimal SUSY and an
extended one (light blue) with potentially suppressed interaction
strengths in non--minimal SUSY models. 

Another example of a WIMP that
would belong to the light blue box is the the lightest Kaluza--Klein
state which is massive, fairly weakly interacting and stable in some
extra--dimensional frameworks~\cite{st02}.  One can see
that a typical strength of WIMP interactions
can be several orders of magnitude less then weak, while still giving
$\abund\sim 0.1$.

Then we have E--WIMPs whose interactions are much weaker than
electroweak.  One well--known example is the axion -- a light neutral
pseudoscalar particle which is a by--product of the Peccei--Quinn
solution to the strong CP problem. Its interaction with ordinary
matter is suppressed by the PQ scale
$\sim(\mw/\fa)^2\sigmaew\sim10^{-18}\sigmaew\sim10^{-20}\pb$
($\fa\sim10^{11}\gev$), hence extremely tiny, while its mass $m_a\sim
\Lambda_{QCD}^2/\fa\sim \left( 10^{-6}-10^{-4}\right)\ev$ which gives
$\Omega_a\sim0.1$. The axion, despite being so light, is of CDM--type
because it is produced by the non--thermal process of misalignment in
the early Universe.

In SUSY, the axion has its fermionic superpartner,
called axino. Its mass is strongly model--dependent but, in contrast
to the neutralino, often not directly determined by the SUSY breaking
scale $\sim1\tev$. Hence the axino could be light and could naturally
be the LSP, thus stable. An earlier study concluded that axinos could
be {\em warm} DM with mass less than $2\kev$~\cite{rtw90}. More
recently~\cite{ckr} it has been pointed out more massive axinos quite naturally
can be also {\em cold} DM as well, as marked in
Fig.~\ref{bigpic:fig}. 

Lastly, there is the gravitino -- the spin--$3/2$ superpartner of the
graviton -- which arises by coupling SUSY to gravity. The gravitino
relic abundance can be of order one~\cite{gravitinoproduction}  
but one has to also worry 
about the so--called gravitino problem: heavier particles  decay
to gravitinos very late, around $10^8\sec$ after the 
Big Bang, and the associated energetic photons and/or hadrons
may cause havoc to Big Bang Nucleosynthesis (BBN) 
products. The problem is not unsurmountable but more
conditions/assumptions need to be satisfied as I will discuss below.
In Fig.~\ref{bigpic:fig} the gravitino is marked in the mass range of
$\kev$ to $\gev$ and gravitational interactions only, although
light gravitinos have actually strongly enhanced couplings via their
spin--$1/2$ goldstino component.

While Fig.~\ref{bigpic:fig} is really about WIMPs which arise in
attractive extensions of the SM, it is worth mentioning another class
of relics, popularized under the name of WIMPzillas, for which there
exist robust production mechanisms (curvature perturbations) in the
early Universe~\cite{wimpzilla:ref}. As the name suggests, they are
thought to be very massive, $\sim10^{13}\gev$ or so. There are no
restrictions on WIMPzilla interactions with ordinary matter, as schematically
depicted in Fig.~\ref{bigpic:fig}.

In summary, the number of {\em well--motivated} WIMP and WIMP--type
candidates for CDM is in the end not so large.  One can add to this
picture other candidates, but the three candidates for the CDM
predicted by SUSY: the neutralino, the axino and the gravitino are
robust. The neutralino is testable in experimental programmes of this
decade (DM searches and the LHC) and is therefore of our primary
interest but we should not forget about other possibilities.

In this talk, we will discuss two intriguing E--WIMPs: the axino and
the gravitino. I will demonstrate many similarities they share as well
as many cosmological and phenomenological differences which may give
one a chance to distinguish them from each other and from the standard
neutralino at the LHC.

\section{The Axino} 

The axino is a  superpartner of the axion.  It is a
neutral, $R=-1$, Majorana, chiral, spin--$1/2$ particle. There exist
several SUSY and supergravity implementations of the well-known
original axion models (KSVZ~\cite{ksvz} and DFSZ~\cite{dfsz}).
(Axion/axino-type supermultiplets also arise in superstring models.)
In studying cosmological properties of axinos, we will concentrate on
KSVZ--type models where the global $U(1)$ PQ symmetry is
spontaneously broken at the PQ scale $\fa$. A combination of
astrophysical (white dwarfs, \etc) and cosmological bounds leads to
$10^9\gev\lsim\fa\sim10^{12}\gev$\cite{axionreviews:cite} although
the upper bound can be significantly relaxed if inflation followed the
decoupling of primordial axionic particles and the reheating
temperature $\treh\ll\fa$.

The two main parameters of interest to us are the axino mass and
coupling. The mass $\maxino$ strongly depends on an underlying model
and can span a wide range, from very small ($\sim\ev$) to large
($\sim\gev$) values. In contrast to the
neutralino (and the gravitino), axino mass does not have to be of the
order of the SUSY breaking scale in the visible sector, $\msusy\sim
100\gev - 1\tev$~\cite{rtw90,ckn}.
%
In a cosmological study of axinos, we will treat axino mass as a free
parameter.
 
Axino couplings to other particles are generically 
suppressed by $1/\fa$.
At high temperatures the most important coupling will be
that of an axino--gluino--gluon dimension--five interaction term
\begin{equation}
i \frac{\alpha_s}{16\pi\left(\fa/N\right)}
{\bar{\axino}}\gamma_5[\gamma^\mu,\gamma^\nu]\gluino^b F^b_{\mu\nu},
\label{eq:axino-gluino-gluon}
\end{equation}
where $\gluino$ stands for the gluino and $N=1(6)$ for the KSVZ (DFSZ)
model. At low temperatures, on the other hand, often a dominant role
will be played by an analogous coupling of axino--photon--neutralino
\begin{equation}
i \frac{\alpha_Y \cayy}{16\pi\left(\fa/N\right)} 
{\bar{\axino}} \gamma_5 [\gamma^\mu,\gamma^\nu]\bino B_{\mu\nu},
\label{eq:axino-bino-photon}
\end{equation}
where $\bino$ denotes the bino, the fermionic partner of the $U(1)_Y$
gauge boson $B$, which is one of the components of the neutralino.
Depending on a model,
there are also terms involving dimension--four operators coming,
\eg, from the {\em effective} superpotential $\Phi\Psi\Psi$ where
$\Psi$ is one of MSSM matter (super)fields.  Axino production
processes coming from such terms will be suppressed at high energies
relative to processes involving Eq.~(\ref{eq:axino-gluino-gluon}) 
and~(\ref{eq:axino-bino-photon}) by
a factor $m_\Psi^2/s$ where $s$ is the square of the center of mass
energy. Such terms will play an important role in axino production
from squark and slepton decays at low temperatures.

\paragraph{Axino Production} 
Because axino interactions are strongly
suppressed, their initial thermal
population decouples at very high temperatures.
We further assume that it (and those of other
relics, like gravitinos) present in the early Universe was
subsequently diluted away by inflation and that $\treh<\Td$. It also
had to be less than $\fa$, otherwise the PQ would have
been restored thus leading to the well-known domain wall problem
associated with global symmetries.

There are two generic ways of repopulating the Universe with axinos.
First, they can be generated through {\it thermal production} (TP),
via scatterings and decay processes of ordinary particles and
sparticles in thermal bath.  Second, they may also be produced in
decay processes of particles which themselves are
out--of--equilibrium, in {\it non thermal production} (NTP).
These production mechanisms will also apply  to the gravitino E--WIMP.

\paragraph{Thermal Production} 
The thermal production can be obtained by integrating the Boltzmann
equation
with both scatterings and decays of particles in the  plasma.

The main axino production channels are the scatterings of
(s)particles described by a dimension--five axino-gaugino-gauge boson
term, Eq.~(\ref{eq:axino-gluino-gluon}).
Because of the
relative strength of $\alphas$, the most important contributions will
come from 2--body strongly interacting processes into final states,
$i+j\ra \axino+\cdots$.

Axinos can also be produced through decays of heavier
superpartners in thermal plasma. At  $T\gsim\mgluino$
these are dominated by the decays of gluinos into LSP axinos and
gluons.  
At lower temperatures $\mchi\lsim T\lsim\mgluino$, neutralino
decays to axinos also contribute while at higher temperatures they are
sub--dominant. 
The TP contribution to axino abundance will be denoted by $\abundatp$.
\begin{figure}[t!]
  \begin{tabular}{c c}
    \includegraphics[width=0.5\textwidth]{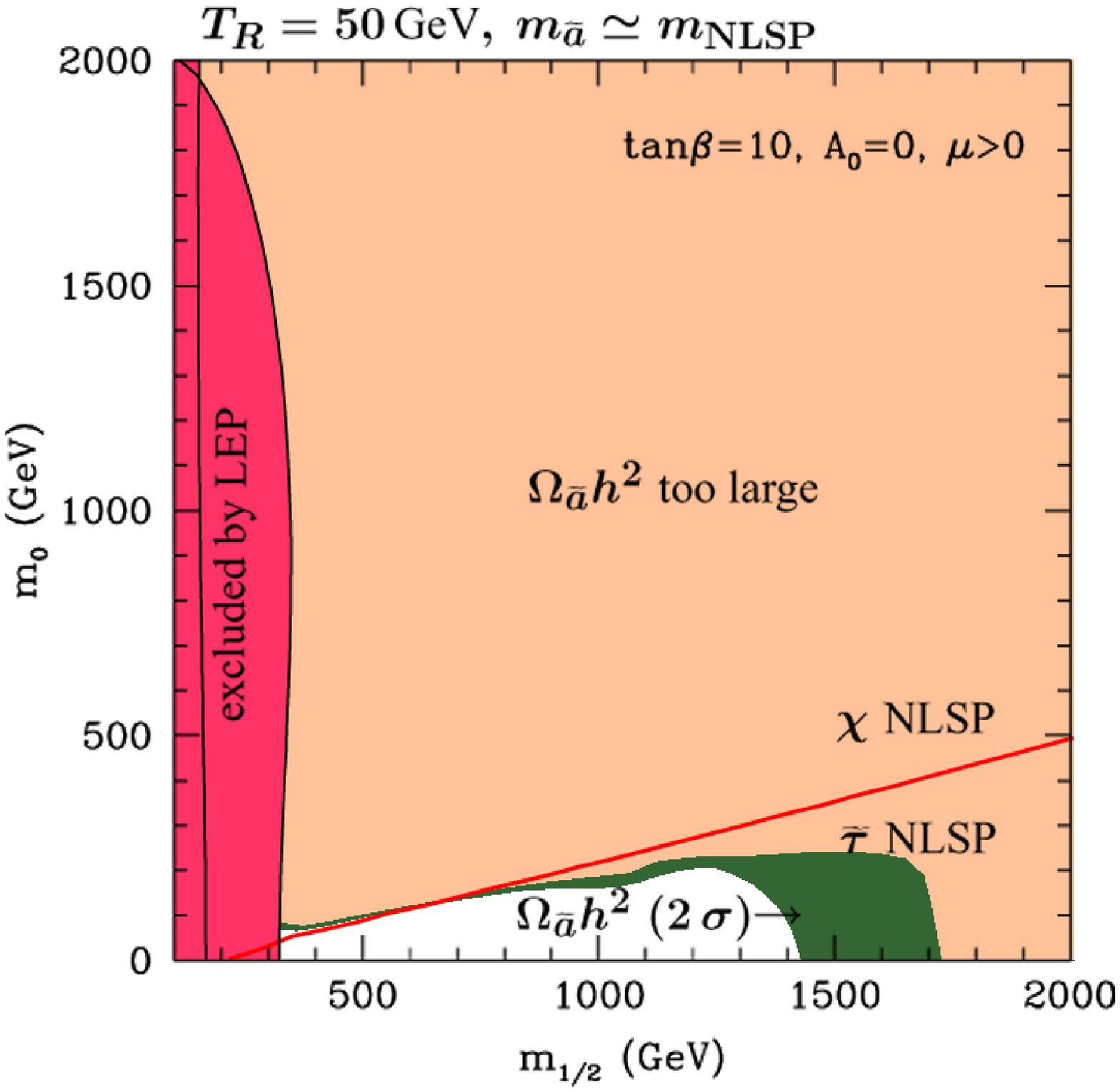}
    & 
    \includegraphics[width=0.5\textwidth]{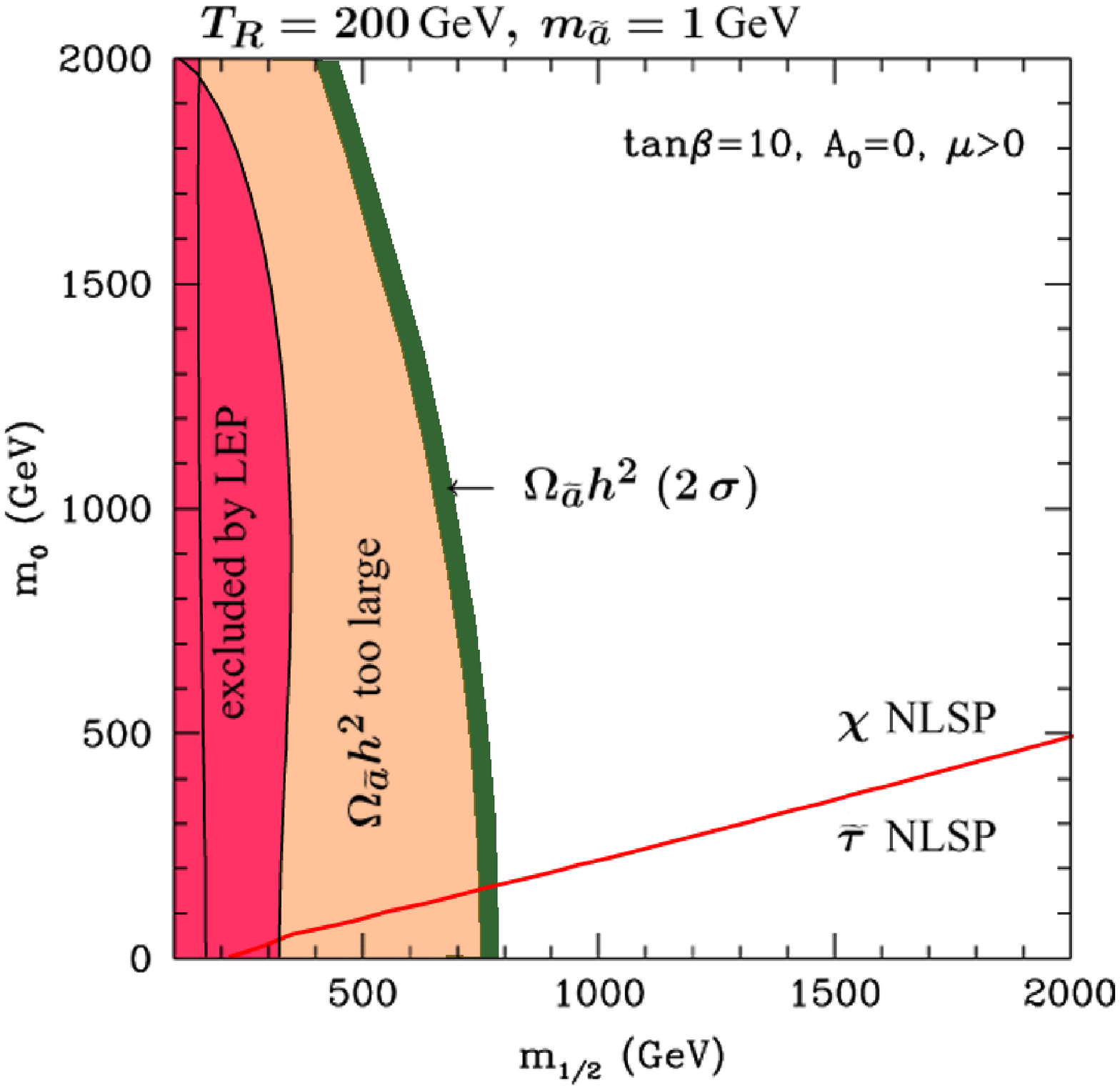}
  \end{tabular}
  \caption{\small The plane ($\mhalf,\mzero$) for axino only slightly
less massive than the NLSP and $\treh=50\gev$ (left window) and for
$\maxino=1\tev$ and $\treh=200\gev$ (right window). We take
$\tanb=10$, $\azero=0$, $\mu>0$ and $\fa=10^{11}\gev$. The regions
excluded by LEP are shown in red. The dark green
(orange, white) regions correspond to $0.094<\abunda<0.129$ ($\abunda$
too large and excluded, too small but otherwise allowed) The red line
divides the neutralino and stau NLSP regions.  
}
  \label{fig:axinotp+ntp}
\end{figure}
%

\paragraph{Non-Thermal Production} 
As the Universe cools down, all heavier SUSY partners will
first cascade decay to the next--to--lightest superpartner (NLSP),
which we denote by $X$.
The NLSPs then freeze out of thermal
equilibrium and subsequently decay into axinos (or gravitinos).

For example, when the NLSP is a nearly pure bino ($X\simeq\bino$), the
decay time is 
approximately given by~\cite{ckr,ckkr}
\begin{equation}
\begin{split}
  \label{eq:lifebinotoaxino}
\tau(X\rightarrow\axino\,\gamma )\simeq
   0.3\sec\left(\frac{100\gev}{m_X}\right)^3\ldots
\end{split}
\end{equation}
which means that the decay will mostly take place before the epoch of
BBN and an energetic photon produced along the axino will not do any
harm to light elements. This is very different from the gravitino case
which is produced long after BBN.

Since all the NLSPs subsequently decay into axinos, 
the axino abundance $\abundantp$ from NLSP decay can be determined 
by the simple relation of mass ratio
and NLSP abundance $\Omega_X$,
\beq
\Omega_{\axino}^{NTP}=\frac{\maxino}{m_X}\Omega_X.
\label{abundgntp:eq}
\eeq
%

\paragraph{Axino LSP in the CMSSM}
From now on we'll concentrate on the Constrained Minimal
Supersymmetric Standard Model (CMSSM)~\cite{kkrw} in which all the
Higgs and superpartner mass spectra are parametrized in
terms of a common gaugino $\mhalf$ and scalar $\mzero$ mass
parameters, a common trilinear parameter $\azero$ (all defined at the
GUT scale), as well as the ratio of the vev's of the two Higgs
doublets $\tanb$. The CMSSM is a reasonable low--energy framework for many
GUT--based models and is also of  experimental interest as a
benchmark model for the LHC. 

In the CMSSM, the NLSP is typically either the (bino--dominated)
neutralino $\chi$ (for $\mhalf\ll \mzero$) or the lighter stau
$\stauone$ (for $\mhalf\gg \mzero$). It can decay to the axino and
photon $\chi\ra\axino\gamma$ or via $\stauone \ra \axino\tau$.


Below we illustrate axino LSP as CDM with two examples taken from
Ref.~\cite{crs,crrs}.  In Fig.~\ref{fig:axinotp+ntp}, the
($\mhalf,\mzero$) plane is shown for $\tanb=10$ and for $\treh=50\gev$,
$m_{\axino}\simeq m_{NLSP}$ (left window) and $\treh=200\gev$,
$m_{\axino}=1\gev$ (right window).  We apply several constraints from
colliders: the lower mass bounds on chargino $ m_{\chi^\pm} >
104\gev$, Higgs $m_h > 114.4\gev$ and stau $m_{\stauone}> 87 \gev$ at
LEP and from
$BR(B\rightarrow X_s\gamma)=(3.34\pm0.68)\times 10^{-4}$.
In addition, the orange colored region is excluded by the overclosure
of Universe while the white regions are cosmologically allowed but not
favored.  The (green) band between the orange and white regions is
the cosmologically favored ($2\,\sigma$) region of
$\abunda=\abundatp+\abundantp$ consistent with observations.  The
recent WMAP results combined with other measurements imply the
$2\,\sigma$ range for non-baryonic CDM
$0.094<\abundcdm < 0.129$~\cite{wmap_cdm}.

In the left window NTP plays a dominant role. We have chosen the axino
mass as large as possible (slightly lighter than NLSP).  Almost all
the neutralino NLSP region is excluded by an over--abundance of CDM
except for tiny region of small $\mhalf$ and $\mzero$. (This case is
nearly identical to the standard case of neutralino LSP.)  A
cosmologically favored region mostly lies in the stau NLSP regime.

In the right window, where we have chosen a small axino mass, the
dominant contribution comes from TP and the overall pattern of
cosmologically excluded and allowed regions is very different.  In
particular, the region closer to the $\mzero$ axis is excluded. As
$\treh$ increases, it grows and pushes away the green band of the
cosmologically favored region farther to the right.  Since, due to the
small $\maxino$, NTP is negligible, the band of $\abunda\sim0.1$
extends to both the neutralino and the stau NLSP regions.

In both windows one finds that the cosmologically favored region lies
in the stau NLSP region (below red line), which is traditionally
believed to be excluded as corresponding to a stable charged relics.

Note also a rather low $T_R$. In fact, this is a characteristic
feature of axino CDM scenario~\cite{ckr,ckkr,crs,crrs} where NTP allows
for $\treh\lsim10^4\gev$ and TP for somewhat larger values $\lsim10^6\gev$.

\section{THE GRAVITINO}

Let us now examine the gravitino as the true LSP and a candidate for
CDM.\footnote{Here we mostly follow Refs.~\cite{rrc04,ccjrr05}.}  Its
mass arises through the super--Higgs mechanism and in a
gravity--mediated SUSY breaking case is naturally expected in the
range $m_{\gravitino} \sim \gev - \tev$.  In other SUSY breaking
scenarios it can be much smaller or larger. As before with the axino,
here we will take the gravitino mass to be a free parameter.


The gravitino can be produced in very much the same way as the axino,
via TP and NTP. One crucial difference is that, relative to
Eqs.~(\ref{eq:axino-gluino-gluon}) and~(\ref{eq:axino-bino-photon}),
in the denominators of analogous dimension--five and four terms there
appears a square of the gravitino mass $\mgravitino$, in addition to
$\fa/N\ra \mplanck$. This leads to a different dependence of the
gravitino yield on $\mgravitino$. In particular, after integrating the
Boltzmann equation, for the TP part one finds~\cite{bbp98,bbb00}
\begin{equation}
\abundgtp\simeq 0.2 \left(\frac{\treh}{10^{10}\gev}\right)
\left(\frac{100\gev}{\mgravitino}\right) 
\left(\frac{\mgluino(\mu)}{1\tev}\right)^2,
\label{eq:abundgbbb}
\end{equation}
where $\mgluino(\mu)$ above is the running gluino mass.  One can see
that, for natural ranges of $\mgluino$ and $\mgravitino$, one can have
$\abundgtp\sim 0.1$ at $\treh$ as high as $10^{9-10}\gev$.

For NTP production, on the other hand, one simply replaces
$\maxino\ra\mgravitino$ in Eq.~(\ref{abundgntp:eq}). 
NLSP first freeze out and then decay into gravitinos at late times
which strongly depend on the NLSP composition and mass $\mx$, and on
the decay products~\cite{feng03-prl,eoss03-grav}. The lifetime is roughly given by
\beq
\taux\sim
10^8\sec\left(\frac{100\gev}{\mx}\right)^5
\left(\frac{\mgravitino}{100\gev}\right)^2
\label{lifetime:eq}
\eeq
for $\mgravitino\ll \mx$.  Thus in the parameter space allowed by
other constraints it can vary from from $\gsim10^8\sec$ at smaller
$\mx$ down to $10^2\sec$, or even less, for large $\mhalf$ and/or
$\mzero$ in the $\tev$ range.  This happens during or after BBN, which
gives a strong constraints on the gravitino LSP scenario.

\begin{figure}[t!]
  \begin{tabular}{c c}
    \includegraphics[width=0.5\textwidth]{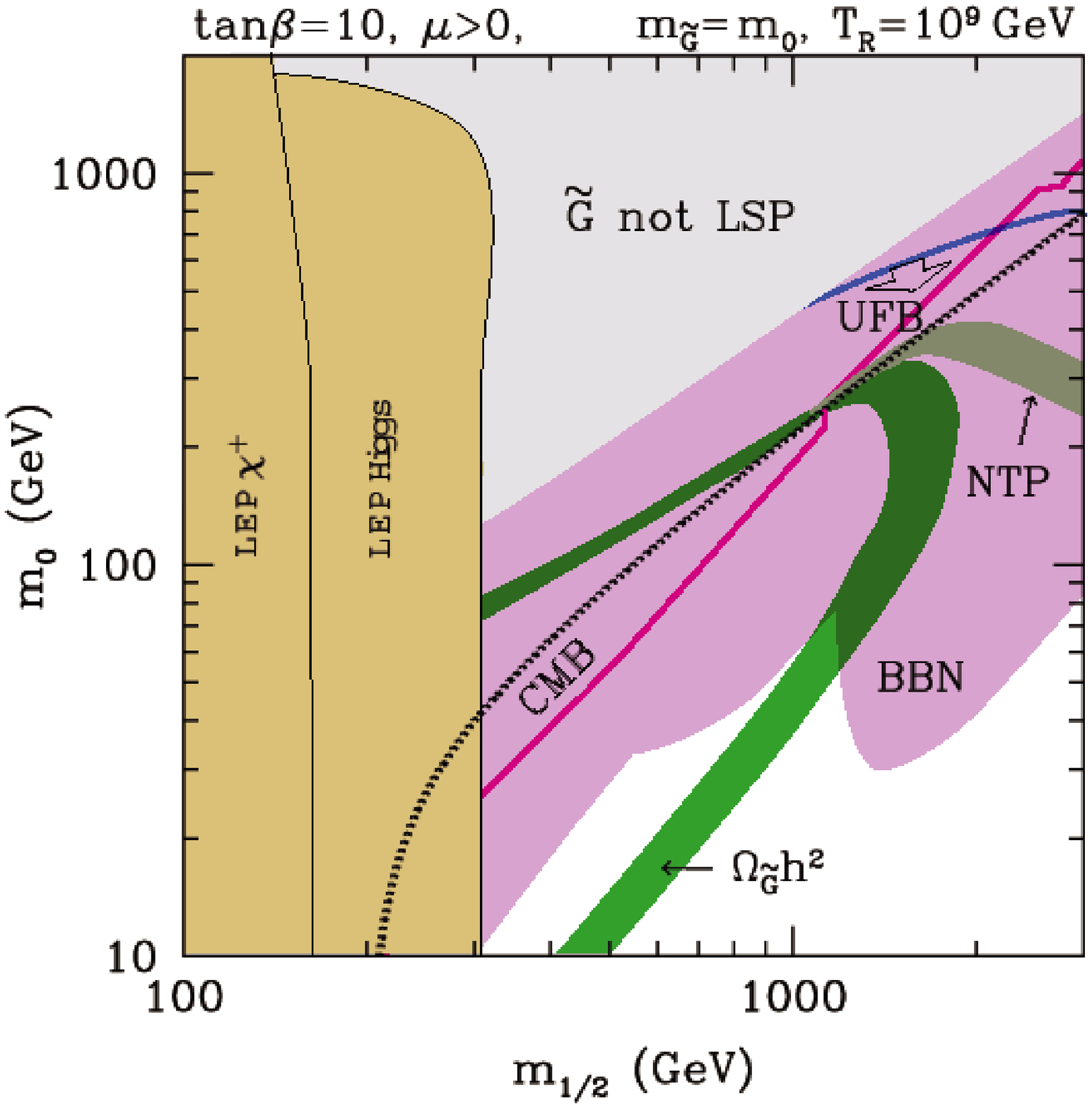}
    & 
    \includegraphics[width=0.5\textwidth]{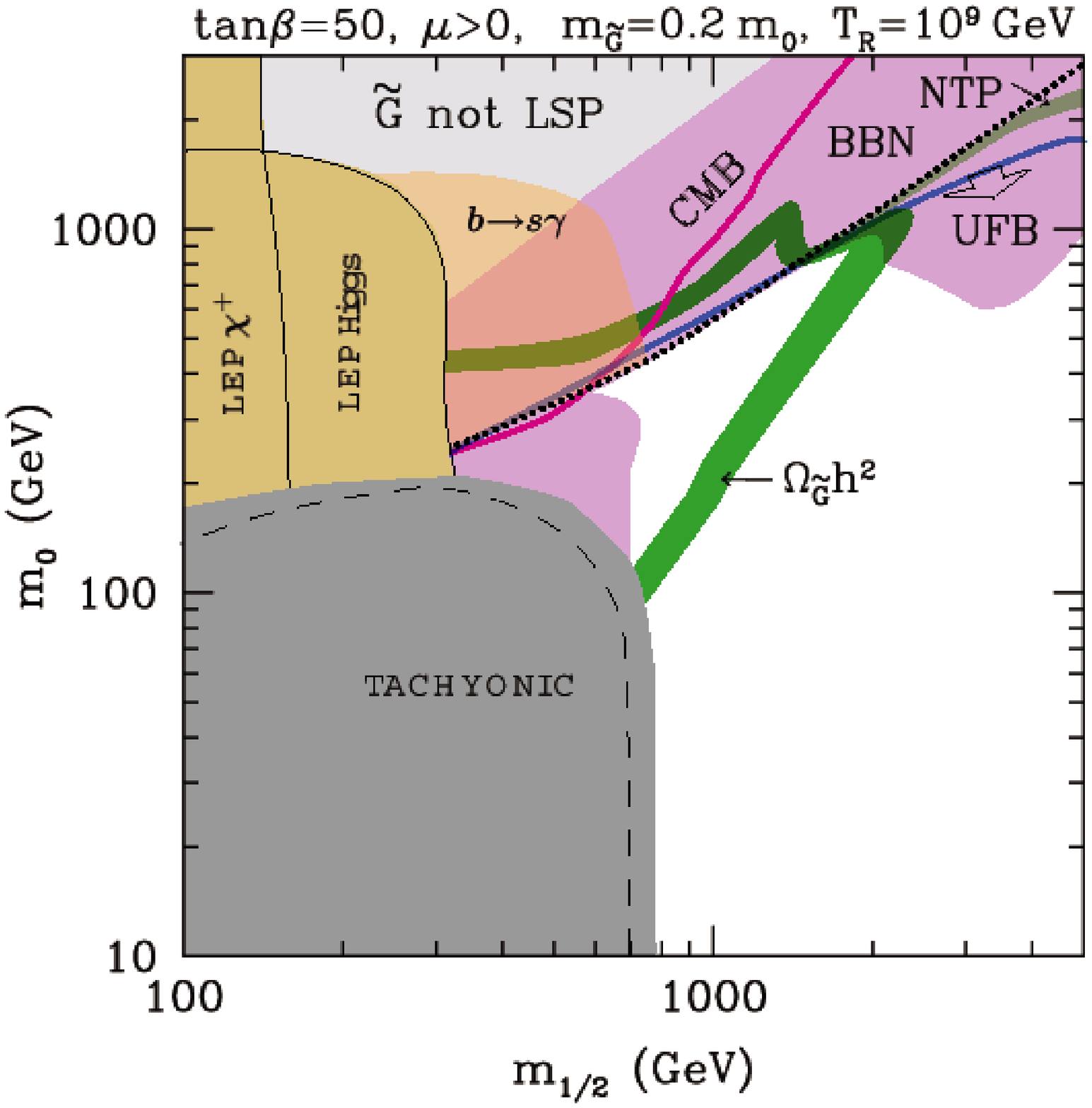}
  \end{tabular}
  \caption{\small The plane ($\mhalf,\mzero$) for $\tanb=10$ (left
window) and $\tanb=50$ (right window) and for $\azero=0$ and
$\mu>0$. The light brown regions labelled ``LEP $\chi^+$'' and ``LEP
Higgs'' are excluded by unsuccessful chargino and Higgs searches at
LEP, respectively. In the right window the darker brown regions
labelled ``$b\to s\gamma$'' is excluded assuming minimal flavor
violation and the dark grey region below the dashed line is labelled
``TACHYONIC'' because of some sfermion masses becoming tachyonic and
is also excluded. In the rest of the grey region (above the dashed
line) the stau mass bound $\mstauone>87\gev$ is violated. In both
windows, the dark green bands labelled ``$\abundg$'' denote the total
relic abundance of the gravitino from both thermal and non--thermal
production is in the favored range, while in the light green regions
(marked ``NTP'') the same is the case for the relic abundance from NTP
processes alone. The regions excluded
by the various BBN constraint are denoted in violet.  
A solid magenta curve labelled ``CMB''
delineates the region (on the side of label) inconsistent with the
CMB spectrum. The UFB  constraints disfavor all the 
stau NLSP region that has remained allowed after applying the BBN
and CMB constraints.}
  \label{fig:bbn+ufb}
\end{figure}

In Fig.~\ref{fig:bbn+ufb} we display the ($\mhalf,\mzero$) plane (now
on a log--log scale) for two representative choices: $\tan\beta=10$
and $m_{\gravitino}=\mzero$ (left window) and $\tan\beta=50$ and
$m_{\gravitino}=0.2 \mzero$ (right window), and for $A_0=0$ and
$\mu>0$.  In addition to the regions excluded by LEP (on the left), we
also reject those where the gravitino is not the LSP or where some
sparticles become tachyonic, as in the right window. The total
gravitino abundance (NTP+TP) consistent with WMAP is shown in dark
green region.  The right and upper region of the band is
cosmologically excluded from the over--abundance of DM.  The light
green region (marked ``NTP'') corresponds to the case when NTP alone
is considered, which is dominant at low $T_R$.

%
%

\paragraph{BBN constraints}
The $\chi$ NLSP decays predominantly to photons via $\chi \ra
\gravitino + \gamma$ and (in a small fraction) to hadrons $\chi\ra\gravitino
\gamma^\ast/Z^\ast \ra \gravitino q\bar q$.  For the $\stauone$ NLSP,
the dominant mode is $\stauone\ra\gravitino\tau$ and hadronic jets are
produced by 4--body decays with suppressed branching ratio.  However
if kinematically allowed, higher hadronic branching ratio can be
obtained through subsequent $Z/$Higgs bosons decays after NLSP decay
to them. 

The decay products, such as photons (tau leptons for $\stau$ NLSP) and
hadrons, carry high energies inherited from the parent NLSP.  These
energetic particles may change the abundances of light elements
produced during the epoch of BBN thus spoiling a good agreement of
light element abundances between predictions and observations.

The injection of high energy photons (or charged leptons) at late 
times ($\tau\gtrsim 10^4 \sec$)  
can disintegrate $D$ (for $\tau \lesssim 10^6 \sec$) 
and $^4He$ (for $\tau \gtrsim 10^6 \sec$) leading to the production
of other lighter elements, such as $D$, ${^3}{\! He}$ or ${^6}{\! Li}$.

At earlier times ($\tau \lesssim 10^4 \sec$) hadronic showers are
induced by mesons for $\tau \lesssim 10^2 \sec$ and nucleons for $\tau
\gtrsim 10^2 \sec$. Mesons convert protons to neutrons resulting in an
increase of $^4He$ abundance, and nucleons increase the abundance of
$D$ and $^6Li$.

We calculate the light element abundances in a self--consistent way
using a code in~\cite{jedamzik04} and compare them with
observations.  Here we adopt the conservative abundances of light
elements,
\begin{center}
\begin{tabular}{r c l}
$2.2\times10^{-5}<$ & $D/H$ & $< 5.3\times10^{-5}$ 
\\ 
$0.232 <$ & $Y_p$ & $< 0.258$ 
\\
$8\times10^{-11}<$ & ${^7}{\! Li}/H$ & 
\\

& ${^3}{\! He}/D$ & $< 1.72$ 
\\
& ${^6}{\! Li}/{^7}{\! Li}$ & $< 0.1875$.
\end{tabular}
\label{lightelements:table} 
\end{center}

In Fig.~\ref{fig:bbn+ufb} the regions excluded by constraints from
light element abundances are shown in violet and marked ``BBN''.  We
can see that the neutralino NLSP region is not viable~\cite{fst04}
while large parts of the stau NLSP domain are, and this is also where the total
gravitino relic abundance in the range consistent with
WMAP.\footnote{For much smaller $\mgravitino\ll 1\gev$, but still
consistent with the gravitino being cold DM, the neutralino region
becomes allowed again.} On the other hand, the (light green) region of
NTP only is not consistent with BBN. Thus a substantial contribution
from TP, and therefore a large enough $\treh$ is required. On the
other hand, increasing $\treh$ causes the green band of $\abundg$ to move
to the left, which eventually runs into conflict with BBN.  In the
moderate gravitino mass region we could find a reheating temperature
as high as $T_R \lesssim 4\times 10^9\gev$ consistent with all other
constraints.

\paragraph{CMB constraints}
The radiative decay process of NLSP releases a net photon energy into the
electromagnetic plasma. For late decays, number changing interactions,
such as thermal bremsstrahlung and double Compton scattering, may be
ineffective, resulting in a discrepancy with Planckian distribution~\cite{hu93}.
The observed Planckian shape of CMB gives the constraints on  the upper 
bound of a dimensionless chemical potential $\mu$,
$|\mu| < 9\times 10^{-5}$~\cite{mubound}.
For later decays ($\tau \gtrsim 4\times 10^{11} \simeq 8.8\times 10^9\sec$)
constraints on CMB can be described by the Compton parameter 
$y$ where
$|y|<1.2\times 10^{-5}$~\cite{pdg02}.

A magenta line in Fig.~\ref{fig:bbn+ufb} delineates the region
inconsistent with CMB spectrum.  We find that the CMB constraints
is usually less stringent than the constraints from the BBN.

\paragraph{False vacuua}
The presence of scalars with color and electric charge in SUSY
theories induce a possible existence of charge and color breaking
(CCB) minima. Also along some directions in field space the potential
can even become unbounded from below (UFB) at tree level. (After
including one loop corrections a UFB direction develops a deep CCB
minimum.)  The most dangerous one is the UFB--3 direction involving
the scalar fields $\{H_u,\nu_{L_i},e_{L_j},e_{R_j}\}$ with $i \neq
j$~\cite{clm1}.  By simple analytical minimization of relevant terms
of the scalar potential and requiring $V_{\rm UFB-3}(Q=\hat Q) >
V_{\rm Fermi}$, where $\hat{Q}$ is the minimization scale and $V_{\rm
Fermi}$ is the Fermi minimum evaluated at the typical scale of SUSY
masses, one obtains a constraint on the SUSY parameter space.

In Fig.~\ref{fig:bbn+ufb}, regions corresponding to our vacuum being a
false vacuum are delineated by a blue line (on the side of a big
arrow) and marked ``UFB''. We can see that the  constraint
disfavors almost all of the stau NLSP region. (Note that this bound is
not specific to the gravitino and applies to the
axino case as well.) However, the existence
of such a dangerous global vacuum cannot be excluded since the color
and electric neutral Fermi vacuum which the Universe is in may be a
long--lived local minimum.  In this case a non-trivial constraints is
placed on the inflationary cosmology~\cite{fors95,forss96}.

\section{Conclusions}
E--WIMPs are well motivated, attractive and intriguing candidates for
CDM.  Here we have presented two cases of the axino and the gravitino
as stable relics and CDM in the Universe. In the CMSSM, while the
neutralino and stau NLSP remain allowed for the axino LSP, for GeV
gravitino LSP the neutralino NLSP is excluded and only the stau NLSP
remains allowed. In the stau region our vacuum corresponds to a local
minimum while in the global one color and/or electric charge are not
conserved. This is one lesson that we can learn should at the LHC a
massive, electrically charged particle (the stau) be observed as an
effectively stable state, rather then the neutralino. If enough of
them were accumulated, one could possibly observe their decays and
very different differential event distributions~\cite{bchrs05} could,
at least in some cases, allow one to decide whether Nature has chosen
the axino or the gravitino as a stable relic and cold dark matter in
the Universe.

\begin{theacknowledgments}
L.R. would like to thank Profs. J.E.~Kim and K.~Choi, and to the PASCOS--05
Local Committee for their kind invitation and for organizing a stimulating
meeting in an attractive location.

\end{theacknowledgments}



\begin{thebibliography}{99}

\bibitem{jkg90} G.~Jungman, M.~Kamionkowski and K.~Griest,
\prep{267}{1996}{195}.  

\bibitem{lrpramana03} For a recent review, see, \eg, 
L.~Roszkowski, 
{\it Pramana} {\bf 62} (2004) 389 [hep-ph/0404052].

\bibitem{lr91} 
L. Roszkowski, 
\plb{262}{1991}{59}.

\bibitem{kkrw}
G.L.~Kane, C.~Kolda, L.~Roszkowski and J.D.~Wells, \prd{49}{1994}{6173}.

\bibitem{leeweinberg77} 
B.W.~Lee and S.~Weinberg, \prl{39}{1977}{165}. 
See also
E.W.~Kolb and M.S.~Turner,
The Early Universe, Addison-Wesley, Redwood City, 1990.

\bibitem{gk90}
K.~Griest and M.~Kamionkowski, \prl{64}{1990}{615}.

\bibitem{st02}
G.~Servant and T.~Tait, 
\npb{650}{2003}{391}  
and 
New J. Phys. {\bf 4} (2002) 99; 

D.~Hooper and G.D.~Kribs, 
\prd{67}{2003}{055003}. 


\bibitem{rtw90}
K.~Rajagopal, M.S.~Turner and F.~Wilczek, \npb{358}{1991}{447}.

\bibitem{ckr}
L.~Covi, J.E.~Kim and L.~Roszkowski, 
\prl{82}{1999}{4180} [arXiv:hep-ph/9905212].

\bibitem{ckkr}
L.~Covi, H.B.~Kim, J.E.~Kim and L.~Roszkowski, 
\jhep{0105}{2001}{033} [arXiv:hep-ph/0101009].

\bibitem{gravitinoproduction}
H. Pagels and J.R. Primack, 
\prl{48}{1982}{223};
S. Weinberg, \prl{48}{1982}{1303}; 
J.~Ellis, A.D.~Linde and D.V.~Nanopoulos, \plb{118}{1982}{59}; 
\plb{443}{1998}{209}; and several more recent papers.

\bibitem{wimpzilla:ref} 
D.J.H.~Chung, E.W.~Kolb and A.~Riotto, \prl{81}{1998}{4048};
V.~Kuzmin and I.~Tkachev, \prd{59}{1999}{123006}.

\bibitem{ksvz}
J.~E.~Kim, \prl{43}{1979}{103};
M.~A.~Shifman, V.~I.~Vainstein and V.~I.~Zakharov, 
\npb{166}{1980}{4933}.

\bibitem{dfsz}
M.~Dine, W.~Fischler and M.~Srednicki, 
\plb{104}{1981}{99};
A.~P.~Zhitnitskii, \sjnp{31}{1980}{260}.

\bibitem{axionreviews:cite}
J.E. Kim, \prep{150}{1}{1987};
M.S. Turner, \prep{197}{67}{1990};
P. Sikivie, \npps{87}{2000}{41}. 
  
\bibitem{ckn}
E.J. Chun, J.E. Kim and H.P. Nilles, \plb{287}{123}{1992}.

\bibitem{crs}
L.~Covi, L.~Roszkowski and M.~Small, \jhep{0207}{2002}{023} [hep-ph/0206119].

\bibitem{crrs}
L.~Covi, L.~Roszkowski, R.~Ruiz de Austri and M.~Small,
\jhep{0406}{2004}{003} [hep-ph/0402240].

\bibitem{wmap_cdm}
D.N.~Spergel, {\it et~al.}, 
{\it Astrophys. J. Suppl.} {\bf 148} (2003) 175. 

\bibitem{rrc04}
L.~Roszkowski,  R.~Ruiz de Austri and K.-Y.~Choi,
\jhep{0508}{2005}{080} [hep-ph/0408227].

\bibitem{ccjrr05}
D.G.~Cerde\~no, K.-Y.~Choi, K.~Jedamzik, L.~Roszkowski and R.~Ruiz de Austri,
hep-ph/0509275.

\bibitem{bbp98} 
M.~Bolz, W.~Buchm\"uller and Pl\"umacher,
\plb{443}{1998}{209} [hep-ph/9809381].

\bibitem{bbb00} 
M.~Bolz, A.~Brandenburg and W.~Buchm\"uller, 
\npb{606}{2001}{518} [hep-ph/0012052].

\bibitem{feng03-prl}
J.L.~Feng, A.~Rajaraman and F.~Takayama, 
\prl{91}{2003}{011302} [arXiv:hep-ph/0302215].

\bibitem{eoss03-grav}
J.~Ellis, K.A.~Olive, Y.~Santoso and V.~Spanos, 
\plb{588}{2004}{7} [hep-ph/0312262].

\bibitem{jedamzik04}
K.~Jedamzik, 
\prd{70}{2004}{063524} [astro-ph/0402344].

\bibitem{fst04} 
J.L.~Feng, S.~Su and  F.~Takayama, 
\prd{70}{2004}{063514} [hep-ph/0404198] and 
\prd{70}{2004}{075019} [hep-ph/0404231]. 

\bibitem{hu93} W.~Hu and J.~Silk, 
\prl{70}{1993}{2661} and 
\prd{48}{1993}{485}.

\bibitem{mubound}
D.J.~Fixsen, \etal,
\apj{473}{1996}{576};
K.~Hagiwara, \etal,  [Particle Data Group],
\prd{66}{2002}{010001}.

\bibitem{pdg02} K.~Hagiwara, \etal, 
\prd{66}{2002}{010001}.

\bibitem{clm1}
J.A.~Casas, A.~Lleyda and C.~Mu\~noz, 
\npb{471}{1996}{3}.

\bibitem{fors95}
T.~Falk, K.A.~Olive, L.~Roszkowski, M.~Srednicki, 
\plb{367}{1996}{183}.

\bibitem{forss96}
T.~Falk, K.A.~Olive, L.~Roszkowski, A.~Singh, M.~Srednicki
\plb{396}{1997}{50}.

\bibitem{bchrs05}
A.~Brandenburg, L.~Covi, K.~Hamaguchi, L.~Roszkowski and F.D.~Steffen,
\plb{617}{2005}{99} [hep-ph/0501287].


\end{thebibliography}
\end{document}